\documentclass[sigconf]{acmart}
\usepackage{pgfplots}
\pgfplotsset{compat=1.18}
\usepackage{enumitem}
\usepackage{graphicx}
\usepackage{subcaption}

\copyrightyear{2025}
\acmYear{2025}
\setcopyright{cc}
\setcctype{by}
\acmConference[ICHEC 2025]{Proceedings of the 2025 International Conference on Human-Engaged Computing}{November 21--23, 2025}{Singapore, Singapore}
\acmBooktitle{Proceedings of the 2025 International Conference on Human-Engaged Computing (ICHEC 2025), November 21--23, 2025, Singapore, Singapore}
\acmDOI{10.1145/3786995.3787016}
\acmISBN{979-8-4007-2234-9/2025/11}

\AtBeginDocument{%
  }
\begin{document}

\title{User-Centered Design of Hyperlocal Communication Platforms: Insights from the Design and Evaluation of KUBO}

\author{Eljohn Evangelista}
\affiliation{%
  \institution{Institute of Computer Science, University of the Philippines Los Baños}
  \city{Los Baños}
  \country{Philippines}
  }
\email{eyevangelista1@up.edu.ph}

\author{Alyssa Cea}
\affiliation{%
  \institution{Institute of Computer Science, University of the Philippines Los Baños}
  \city{Los Baños}
  \country{Philippines}
  }
\email{afcea@up.edu.ph}

\author{Axel Balitaan}
\affiliation{%
  \institution{Institute of Computer Science, University of the Philippines Los Baños}
  \city{Los Baños}
  \country{Philippines}
  }
\email{aobalitaan@up.edu.ph}

\author{Clark Vince Diala}
\affiliation{%
  \institution{Institute of Computer Science, University of the Philippines Los Baños}
  \city{Los Baños}
  \country{Philippines}
  }
\email{cdiala@up.edu.ph}

\author{Jamlech Iram Gojo Cruz}
\affiliation{%
  \institution{Institute of Computer Science, University of the Philippines Los Baños}
  \city{Los Baños}
  \country{Philippines}
  }
\email{jngojocruz@up.edu.ph}

\begin{abstract}
Effective hyperlocal communication is critical in the Philippines, where delayed or algorithm-filtered updates can leave residents uninformed about emergency advisories and community events. We conducted a user-centered study consisting of contextual inquiry and semi-structured interviews to identify four key barriers: delayed alerts, algorithm-driven noise, language gaps, and digital divides. Guided by these insights, we designed KUBO (\textit{Kumunidad at Balitang Opisyal}), a prototype that integrates a home module for verified local government unit advisories and curated headlines, and a community module for resident-powered neighborhood reports and discussions. Using a within-subjects evaluation design, KUBO significantly reduced task completion times (p-value < 0.001), improved information recall on post-task quizzes (p-value = 0.010), and yielded higher user satisfaction ratings for ease of use, overall satisfaction, and perceived effectiveness compared to Facebook, the commonly used communication platform in the Philippines. These results demonstrate that a dual-channel, inclusive platform can substantially enhance real-time information access, comprehension, and civic engagement in hyperlocal settings.
\end{abstract}

\ccsdesc[500]{Human-centered computing~User studies}
\ccsdesc[300]{Human-centered computing~Empirical studies in HCI}
\ccsdesc[300]{Human-centered computing~Ubiquitous and mobile computing systems and tools}
\ccsdesc[100]{Human-centered computing~Accessibility design and evaluation methods}

\keywords{human-computer interaction, prototyping, information retrieval, informedness, hyperlocal communication}

\maketitle

\section{Introduction}
Communities rely on timely and relevant information about local events and announcements to stay safe and engaged. In many Philippine barangays (the smallest level of government) and campus neighborhoods, mainstream media often lack detailed coverage of local issues. In contrast, social media algorithms often bury urgent announcements beneath popular content. The language barriers and intermittent connectivity further exclude residents, and it forces them to depend on slow word-of-mouth networks.

Existing research highlights both the potential and limitations of current platforms. Social media applications like Facebook and Viber have become central to community communication in the Philippines due to their widespread use and familiarity; however, they are also prone to misinformation, algorithmic bias, and limited accessibility in rural or marginalized areas. In light of these, user-centered design (UCD) studies in public communication tools have demonstrated the value of involving users early to ensure trust, usability, and accessibility. These insights highlight the need for dedicated hyperlocal communication systems that prioritize urgent information, inclusivity, and verified sources.

To address these gaps, we introduce \textbf{KUBO} (\textit{Kumunidad at Balitang Opisyal}, or Community and Official News), a prototype hyperlocal communication platform. \textit{Kubo}, in Filipino, refers to a traditional bamboo hut that symbolizes community life and \textit{bayanihan} (collective action). The name reflects two design inspirations: (1) a Community module, which supports resident reports and discussions aligned with the bayanihan values, and (2) a Home module that aggregates verified advisories and curated headlines to offer a safe and trusted space for official information. KUBO is designed to bridge the digital divide and provide inclusive, reliable, and real-time communication.

The significance of this study is the design and evaluation of a prototype system tailored for hyperlocal Philippine communities. Reliable communication has never been more critical, not only for disaster response and public safety, but also for everyday civic engagement and community building. KUBO aims to address shortcomings in existing tools by drawing on cultural values of trust and cooperation, and combining verified information with resident participation.

Specifically, the objectives of this study are: 
\begin{enumerate}
    \item to conduct user research through contextual inquiry and interviews to understand communication needs;
    \item to apply user-centered design methods in informing the design of a hyperlocal communication application;
    \item to iteratively design and prototype KUBO based on identified user requirements; and
    \item to evaluate the prototype against Facebook in a controlled study, focusing on task completion time, informedness, and user satisfaction.
\end{enumerate}

To ground KUBO in context, we next review community communication practices in the Philippines, the challenges of misinformation and algorithmic bias, and the user-centered approaches to address these gaps.

\section{Background and Related Work}

\subsection{Communication Channels and Practices in the Philippines}

Local government units (LGUs) in the Philippines have long used a combination of traditional and digital communication channels to reach constituents, reflecting both the strengths of widespread media adoption and persistent gaps in equitable information access. Historically, LGUs relied on traditional media like radio and television broadcasts, press releases, public forums, town hall meetings, and printed materials, which are particularly vital in rural areas with limited internet connectivity. The Philippine Information Agency (PIA), the official public information arm, collaborates with local networks to disseminate government programs and services \cite{pia2020local}. However, the rapid adoption of digital technologies has fundamentally reshaped community communication. Filipinos spend an average of 3 hours and 32 minutes per day on social media \cite{meltwater2024social} which makes platforms like Facebook a dominant channel for hyperlocal government communication. LGUs have also adapted different strategies to disseminate verified government news, such as by establishing official Facebook pages, adopting messaging platforms like Viber and Telegram, and utilizing the Philippine News Agency's (PNA) web-based newswire service.

Despite this multifaceted communication infrastructure, there are significant structural limitations that persist. For instance, a study by Afable \cite{afable2020facebook} analyzing the public Facebook pages of 17 Metro Manila LGUs found that these pages primarily function as one-way information channels and focuses only on official activities and announcements rather than sustained dialogue. While city-level Facebook groups and messaging platforms enable community interaction, their effectiveness is compromised by the absence of structured official content, inconsistent moderation, and vulnerability to misinformation. In addition, closed platforms like Viber and Telegram, even though they are valuable for targeted communication among specific groups, have remained inaccessible to the broader public and dependent on volunteer-based management. These practices reveal that the current approaches are limited by either structural fragmentation, one-way information flow, or exposure to unverified content.

\subsection{Existing Platforms and Their Limitations}

Several digital platforms have emerged to enhance local government communication but remain fragmented and limited in addressing both accessibility and credibility needs. For instance, the eGov PH Super App at the national level focuses on transactions and government service delivery rather than information exchange or community engagement \cite{egovph2022app,pco2025egovph}. City- and barangay-level applications also demonstrate this transaction-focused approach: 1Namayan in Mandaluyong and MyNaga in Naga City enable residents to file service requests and access information, but lack participatory features, verified news sections, SMS integration, or AI-powered content tools \cite{mandaluyong2022namayan,naga2025mynaga}. Similarly, Marikina e-Concern and Baguio in My Pocket serve as one-way channels for issue reporting and e-services without fostering dialogue or community-aware information sharing.

There are alert-based systems such as 1Hope and Alerto PH that provide disaster notifications and emergency reporting capabilities \cite{playgoogle2025hope,dost2025alerto}. The Free Mobile Disaster Alerts Act (RA 10639) mandates the NDRRMC's cell broadcast system for mass emergency warnings. These initiatives improve rapid-response communication but they address emergency needs only and remain isolated from everyday civic updates and sustained community engagement. Globally, comparable platforms reveal similar gaps: hyperlocal social networks like Jodel and Neighbrsnook lack government verification and remain vulnerable to misinformation \cite{playgoogle2025jodel,neighbrsnook2025benefits}. In contrast, citizen-government platforms like Qlue in Jakarta emphasize issue reporting over news dissemination or civic discourse \cite{firdaus2016qlue}. AI-powered news aggregators such as Particle and LetMeKnow deliver summaries of national and global news but do not cater to localized or government-authenticated content \cite{theverge2024particle,playgoogle2024letmeknow}.

Despite the availability of apps specifically designed for credible news and information sharing, Filipinos overwhelmingly continue to rely on Facebook and other social media platforms as their primary sources of news, with recent surveys showing that 94.7\% of Filipinos use Facebook \cite{meltwater2024social,statista2025socmed,spiralytics2025facts}.

\subsection{Misinformation, Algorithmic Bias, and Design Strategies}

The growing reliance on social media platforms for news and updates in the Philippines has created two challenges: misinformation and algorithmic bias, which together threaten the reliability and visibility of critical hyperlocal information. The Philippines is severely affected by information disorder, which Wardle and Derakhshan \cite{wardle2017information} categorize as disinformation (deliberately false information), misinformation (unintentionally false information), and malinformation (genuine information shared with the intention to cause harm). All three types rise in the Philippines during critical events, from electoral disinformation campaigns to COVID-19 vaccine misinformation and false emergency alerts \cite{delacruz2021surfing, beltran2022tsunami, rappler2023disaster}. Filipinos employ information-verification strategies such as source checking and cross-referencing \cite{fabella2022fakenews} to aid these problems but these efforts prove insufficient against pervasive misinformation.

Algorithmic bias worsens these challenges. Platforms like Facebook and TikTok prioritize engagement over accuracy and uses algorithms that rank content by likes, comments, and shares to maximize user retention and advertising revenue \cite{metzler2024social}. This engagement-driven approach creates filter bubbles: a personalized information ecosystems that isolate users from diverse perspectives \cite{bakshy2015exposure}. These filter bubbles prevent residents from discovering official announcements or critical local news unless they actively seek them out \cite{pariser2011filterbubble}. Sensational and fake news also generate more engagement than factual updates, thereby amplifying misinformation and rendering verified, hyperlocal information invisible.

To address these interconnected problems, researchers have identified several design-based interventions. Global studies demonstrate that fact-checking interventions improve factual accuracy across diverse populations \cite{porter2021global}. Additionally, algorithm-free news feeds and community-focused platforms offer promising alternatives. The SHEDR framework utilizes deep learning to detect hyperlocal events while bypassing engagement-based algorithms \cite{hu2021shedr}. Meanwhile, tools like the Bloom plugin enable geotag-based community news dissemination without algorithmic bias \cite{meese2019bloom}. These are some examples of user-centered design research and it shows that credibility indicators, transparency mechanisms, and warning labels, when designed through human-centered approaches, help users identify misinformation \cite{pennycook2021shifting,bhuiyan2021nudgecred}.

\subsection{User-Centered Design for Inclusive Hyperlocal Communication}

User-centered design (UCD) provides a framework for creating trustworthy, usable, and accessible communication tools: qualities often compromised in algorithm-driven social media ecosystems. Unlike commercial platforms that prioritize engagement over relevance, UCD emphasizes designing systems around diverse user needs through iterative participation \cite{hartwig2024landscape}. Research demonstrates that involving users throughout the design process leads to a more effective and contextually appropriate outcomes, especially for misinformation interventions \cite{malhotra2023user}. UCD principles have been successfully applied to hyperlocal platforms. For example, Kavanaugh et al. \cite{kavanaugh2012hyper} showed how user-centered news aggregation enhances civic engagement, while Gulyas and Hess \cite{gulyas2024three} identified community, commitment, and continuity as essential elements of effective local journalism. Mobile applications developed through UCD can improve accessibility by incorporating offline access and SMS support for users with low connectivity, and even localized languages \cite{stephens2018risk, pedrosa2022user}. Trust mechanisms, such as verification badges and transparent sourcing, can help counter misinformation \cite{noman2024designing, gurgun2024motivated}. However, the Philippine context lacks integrated platforms combining verified news, community engagement, SMS accessibility, and AI-powered summarization within a single design.

Our design process applied the Double Diamond model \cite{designcouncil2005double} and Design Thinking methodology \cite{brown2009change}, frameworks that structure design into iterative cycles of discovery, definition, development, and delivery. These complementary approaches guided our systematic effort to address the gaps discussed in the literature: the limited access to localized, real-time community information and the challenges of misinformation in hyperlocal settings.

\section{Design Process}

Following the established frameworks in design, the Double Diamond model and the Design, our four-stage process as illustrated in Figure 1, maps directly onto these frameworks.

\begin{figure*}[!htbp]
    \centering
    \includegraphics[width=0.80\textwidth]{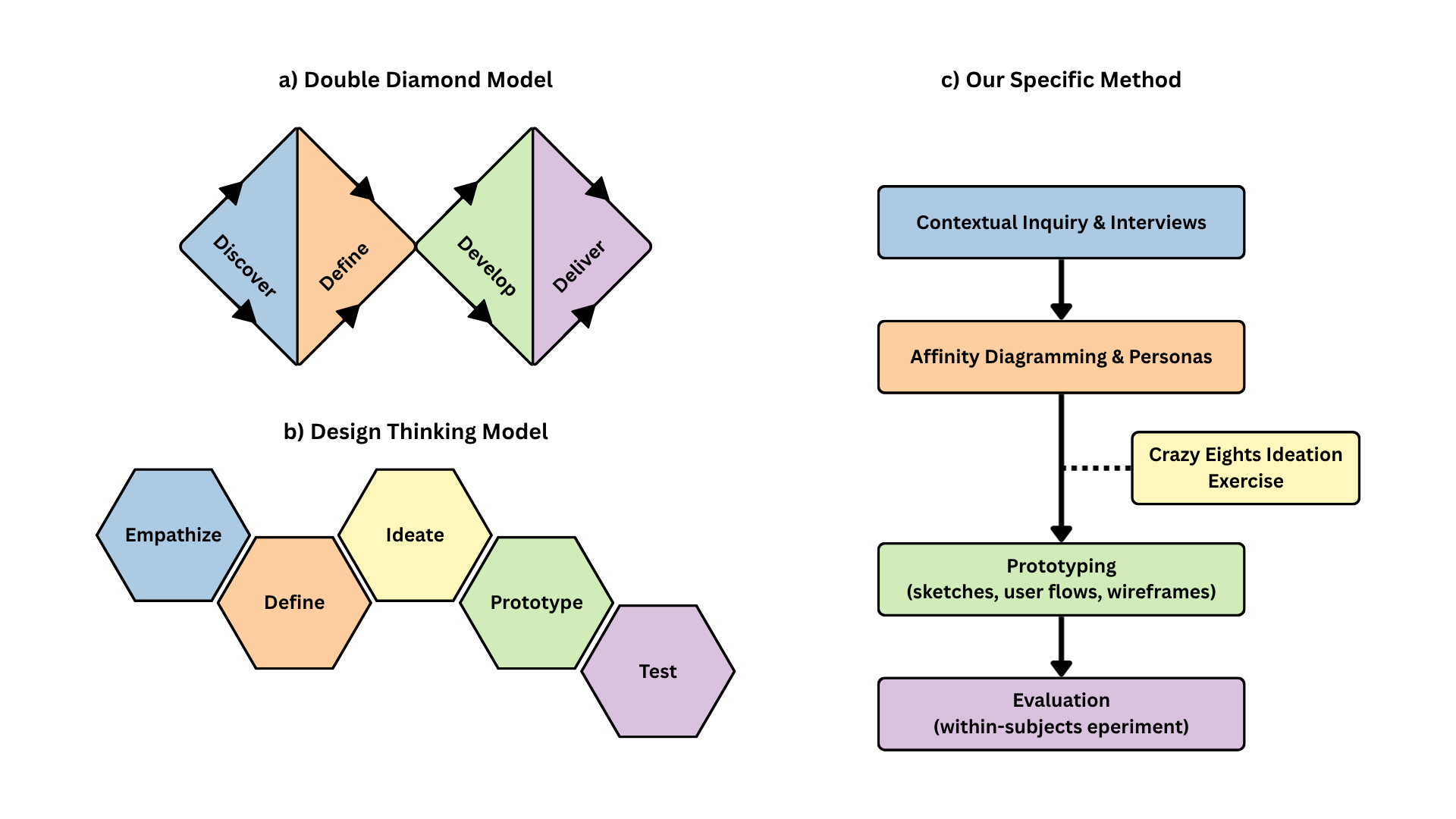}
    \caption{Alignment of our method with established design frameworks.}
    \label{fig:methods-flowchart}
    \Description[Method–framework alignment diagram]{A schematic showing three panes:
    (1) the Double Diamond with Discover→Define→Develop→Deliver,
    (2) the Design Thinking loop with Empathize→Define→Ideate→Prototype→Test,
    and (3) our method mapped onto those stages, highlighting iteration and feedback.}
\end{figure*}

The first two stages addressed the Discover and Define phases, where we conducted contextual inquiry \cite{holtzblatt1997contextual} and semi-structured interviews to understand residents' lived experiences and information access barriers, while the conceptual models like affinity diagramming and persona development \cite{lucero2015using,nielsen2013personas} synthesized the findings into actionable user requirements. The third stage, prototyping, moved through the Develop and Ideate phases by developing iterative prototypes from low- to high-fidelity, supported by user flow diagrams to ensure coherent user experiences. This approach enabled rapid concept testing before investing in detailed interaction design.

The final stage, evaluation, aligned with the Deliver and Test phases through a controlled within-subjects experimental design. We directly compared user experiences and quantified improvements in accessing hyperlocal information, assessing efficiency, informedness, and user satisfaction by having participants interact with both the prototype and Facebook.

Importantly, while presented as a linear sequence, our process was fundamentally iterative, with evaluation and user feedback occurring continuously throughout all stages. This iterative nature reflects both the cyclical structure of the Double Diamond where insights from later phases often necessitate revisiting earlier phases and the non-linear character of Design Thinking, which encourages teams to move fluidly between stages based on emerging insights. 

The subsequent sections detail each major phase: Section 3.1 discuss the user research and data synthesis methods, Section 4 describes the prototyping activities, while Section 5 presents the experimental study used for evaluation.

\subsection{User Research and Data Synthesis}

\subsubsection{Contextual Inquiry and Interviews} To ground our design in actual user needs, we conducted a contextual inquiry and nine semi-structured interviews. While our broader target audience included residents in general, all the participants we interviewed were students from the University of the Philippines Los Baños. 

The contextual inquiry was conducted in the participants' everyday environments, such as outdoor dormitory areas and food establishments they frequented. Most interviews took place in school areas such as the university library and building hallways, and additional sessions were conducted both onsite and online. Each interview lasted around 30 minutes and was transcribed after obtaining informed consent. 

The data collected from these activities provided the foundation for the contextual design models that guided our subsequent analysis.

\subsubsection{Affinity Diagram}
We developed an affinity diagram to synthesize our interview data. This step involved reviewing interview notes and key statements from participants, resulting in 81 unique notes. We grouped similar responses based on recurring themes, which formed 12 mid-level categories such as reliability and credibility, verification of news, time constraints, preference for relevant news, and reliance on friends and family for updates. These categories were then consolidated into five high-level themes: 
\begin{enumerate}
    \item verifying information
    \item preferring mobile and anonymous access
    \item lacking time and facing accessibility issues
    \item getting news from multiple sources
    \item wanting relevant alerts
\end{enumerate}
 
The affinity diagramming process produced five recurring themes that capture the core communication pain points and goals of users. Verifying information was consistently raised as a concern, with residents noting the difficulty of distinguishing legitimate advisories from rumors and unverified social media posts. Participants also emphasized the importance of mobile-first and sometimes anonymous access, which reflects both the reliance on smartphones as the primary medium for connectivity and the need to share or receive updates without fear of judgment. A recurring challenge also involved lacking time and facing accessibility issues: users described being too busy to actively search for updates and often struggling to locate relevant information across cluttered feeds or multiple pages, especially under conditions of low connectivity. Participants also reported having to consult multiple channels to piece together local news, since many community-level updates, such as barangay announcements or neighbor reports, do not reach higher offices or mainstream outlets. This emphasizes the importance of community reporting, but the users also highlighted the need for verification to prevent misinformation. Residents also expressed a strong desire for relevant and timely alerts, particularly for urgent matters such as outages, water interruptions, or safety warnings, which they felt were frequently delayed or hard to find in existing platforms.

\subsubsection{Personas}
\begin{figure}[!htbp]
    \centering
    \includegraphics[width=\columnwidth]{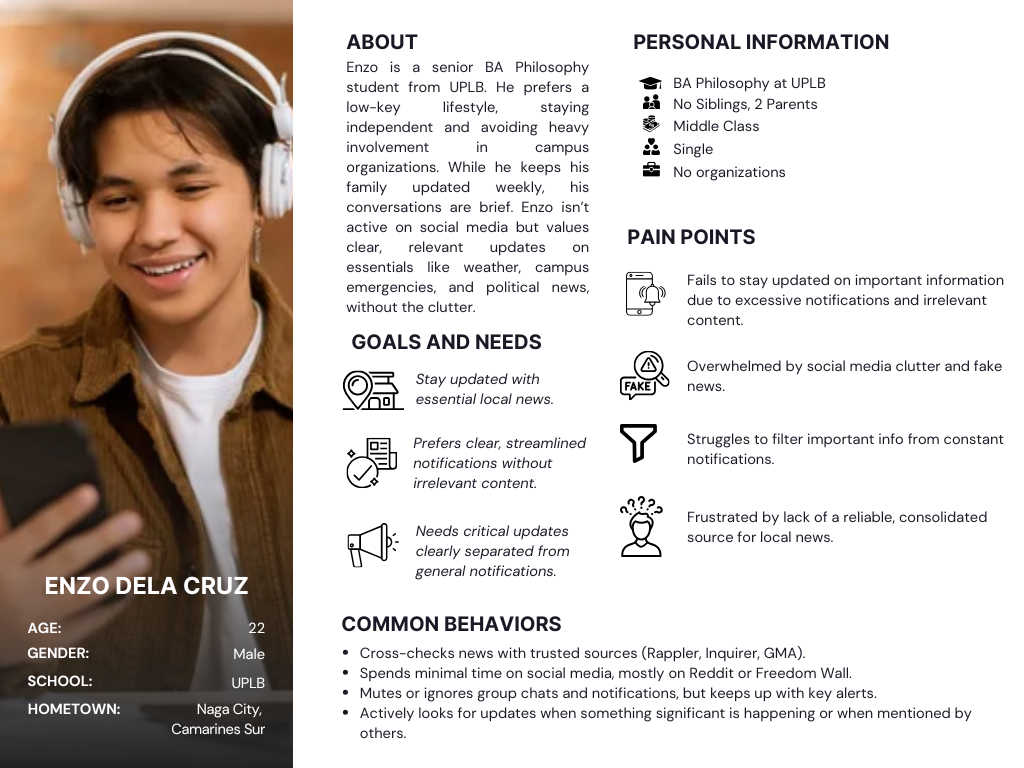}
    \caption{Representative persona generated from contextual inquiries and interviews.}
    \Description[Sample user persona]{A visual profile of a representative community member (i.e., Enzo) showing demographics, goals and needs, pain points, and common behaviors synthesized from contextual inquiries and interviews.}
    \label{fig:persona}
\end{figure}

Building on the affinity diagram, we created personas that reflected the experiences and challenges of our target users. A representative of the four personas developed is shown in Figure~\ref{fig:persona}.

\begin{figure*}[!htbp]
    \centering
    \includegraphics[width=0.8\textwidth]{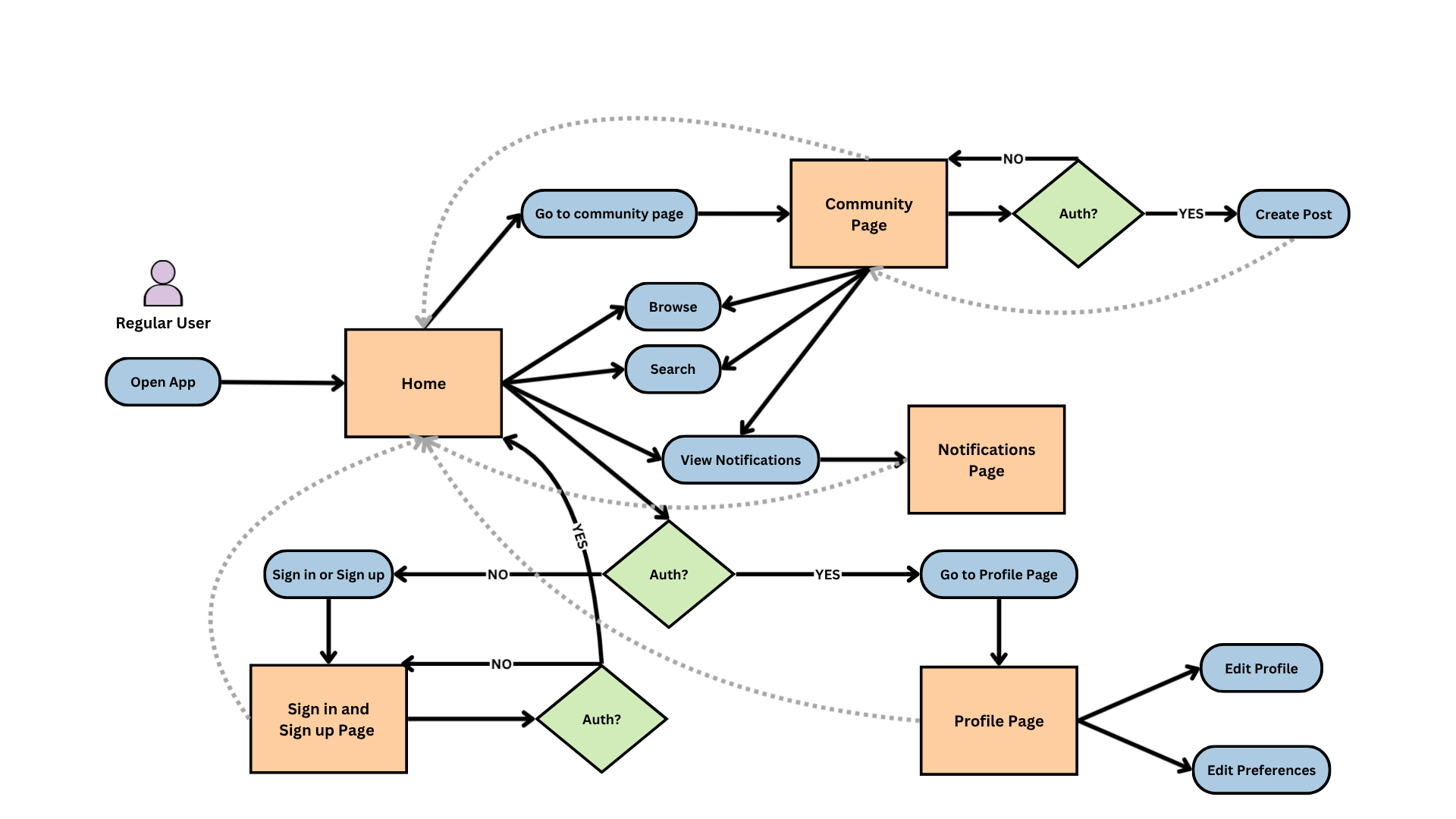}
    \caption{Initial user flow diagram for the KUBO prototype.}
    \Description[KUBO user flow diagram]{A flow diagram showing how users interact with the KUBO hyperlocal communication platform. The flow begins with community members accessing the home screen, followed by options to view verified updates, submit reports, and engage with local posts. Arrows indicate transitions between steps such as authentication, content verification by moderators, and dissemination of approved updates to the community feed.}
    \label{fig:user-flow}
\end{figure*}

Each persona was synthesized from multiple interviewees, often combining one or two primary references with shared behaviors, pain points, and goals across participants. We focused particularly on issues of information access and communication practices.  Each persona was designed to highlight distinct needs and traits. These personas gave us a richer understanding of our users and helped anchor our subsequent design decisions in concrete user experiences.

The conceptualization of the prototype followed the four user personas (i.e., Mia, Risa, Nathan, and Enzo) that represented the diverse realities of hyperlocal communication among students and young professionals. Mia needs accessible and timely academic updates despite limited connectivity to ensure equitable access across varying internet conditions. Risa prefered organized, real-time updates, and credible information to minimize distractions from irrelevant content. Nathan wants to rely on digital communication and desire for social connection while living away from home. Meanwhile, Enzo has frustration with information overload and fake news. Together, these collective and recurring pain points from the personas grounded the prototype's user-centered design and the features for the app.

\begin{figure*}[t]
    \centering
    \begin{minipage}{0.8\textwidth} 
        \centering
        \begin{subfigure}[b]{0.3\textwidth}
            \includegraphics[width=\linewidth]{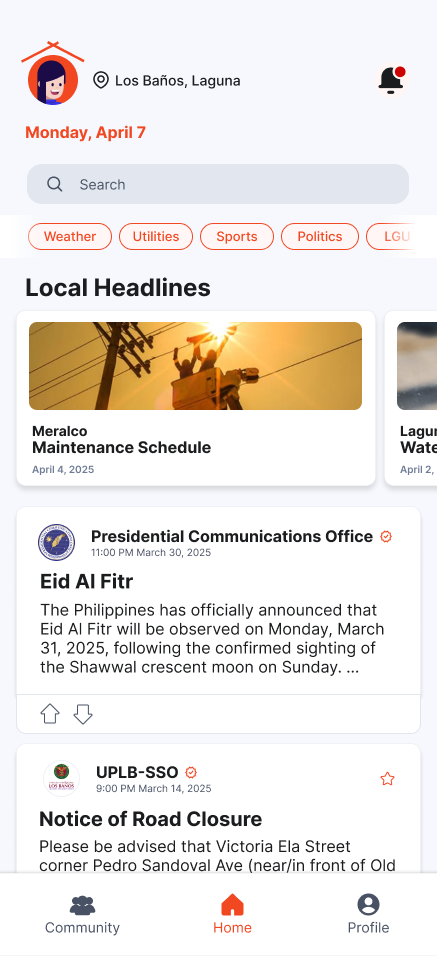}
            \caption{Home with local headlines}
        \end{subfigure}
        \hfill
        \begin{subfigure}[b]{0.3\textwidth}
            \includegraphics[width=\linewidth]{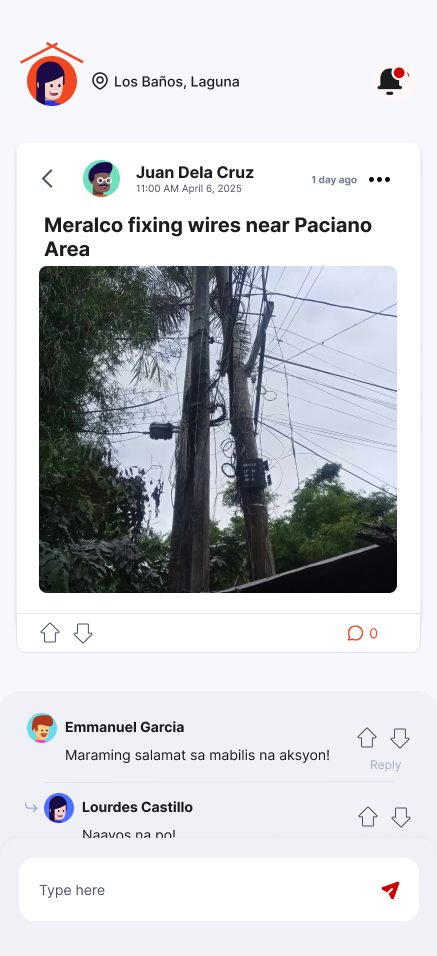}
            \caption{A post in the Community page}
        \end{subfigure}
        \hfill
        \begin{subfigure}[b]{0.3\textwidth}
            \includegraphics[width=\linewidth]{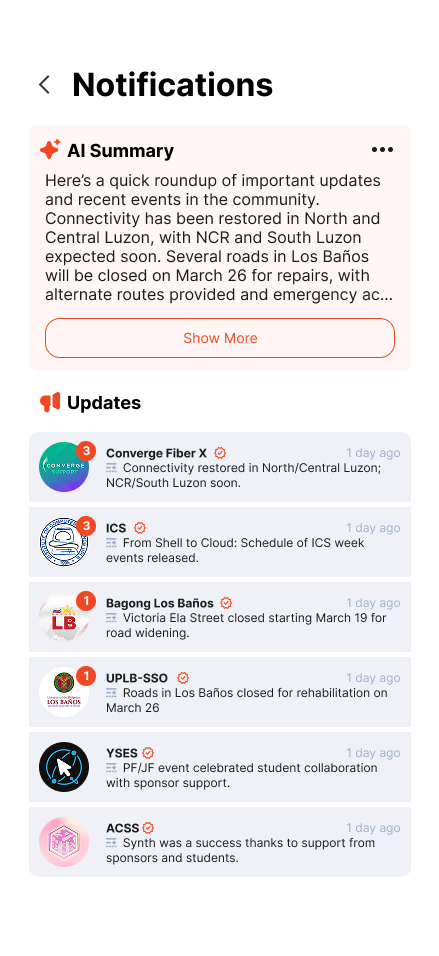}
            \caption{Notifications page}
        \end{subfigure}
    \end{minipage}

    \caption{Screenshots of the KUBO prototype illustrating the main user interface components.}
    \Description[KUBO interface screenshots]{Three screenshots of the KUBO prototype.
    The first shows the Home page with verified local headlines and featured posts. 
    The second displays a single post in the Community page, highlighting engagement features like comments and upvotes.
    The third shows the Notifications page, where users receive alerts about verified updates and responses to their posts, with AI-summarization feature.}
    \label{fig:threephotos}
\end{figure*}

\section{Prototype Overview}

\subsection{Iterative Prototyping}
We developed the KUBO (Komunidad at Balitang Opisyal) prototype through an iterative, user-centered design process. The ideation phase began with the Crazy Eight exercise and this produced early sketches of ideas and possible solutions. In this brainstorming activity, we considered key findings from interviews, contextual inquiries, affinity diagrams, and personas. Afterwards, we identified recurring solutions that informed KUBO's core features:
\begin{enumerate}
    \item a dedicated page for verified news and official posts,
    \item a community feature that fosters interaction and support, and 
    \item SMS alerts for urgent announcements,
    \item integrated AI to summarize long information.
\end{enumerate}

We made an initial evaluation of the ideas based on the needs of the user personas to come up with those primary features. We then translated these ideas into paper sketches and eventually resulted into six basic screens: Login, Signup, Home, Profile, Community, and Notifications. At this stage, we also created a user flow diagram to visualize how these screens would connect, as shown in Figure~\ref{fig:user-flow}. The low-fidelity prototype was intentionally minimal and static with the intention to clarify core functionality while leaving room to refine scope and prioritize features. 

We then developed a medium-fidelity prototype through wireframes using Figma. We decided to maintain a broad scope that addressed the pain points identified during the empathize phase. Core features were solidified in this stage, such as AI-generated summaries for notifications, commenting restricted to the community page, verification badges for official pages, and SMS notifications for urgent alerts. We conducted a heuristic evaluation method \cite{nielsen1995conduct} from potential users (n=18) and feedback gathered at this stage largely validated these design choices. It also highlighted areas for improvement such as refining button UI for clarity, and introducing tags and filters to help users sort through information more effectively.

The high-fidelity prototype incorporated visual polish and additional interactivity features. We refined the interface and added filtering, the ability to follow specific pages, and reporting functions to personalize user experience. These enhancements directly responded to frustrations expressed by participants in earlier phases. For example, the AI summary function addressed information overload, helping users who lacked time to parse long posts. Verification badges for official pages reinforced trust and credibility, as participants noted their need to fact-check social media content. SMS notifications ensured that users could still receive critical updates when offline, while also reducing clutter by reserving SMS for essential alerts. Some screenshots of the prototype is shown in Figure~\ref{fig:threephotos}.

\subsection{Description of Features}

KUBO incorporates four complementary technical features designed to address hyperlocal communication challenges, as mentioned earlier. However, the prototype employs a Wizard-of-Oz approach \cite{dow2005wizard}, a well-established research methodology in human-computer interaction where system behaviors are simulated by a human operator rather than fully automated. Specifically in this study, the AI summarization, SMS alerting, and content verification functions are currently simulated to assess user perception and usability. In this section, we describe how each feature would function upon full technical development.

First, the AI-powered summarization would employ Retrieval-Augmented Generation (RAG) techniques and prompt engineering to reduce hallucinations and maintain factual accuracy and relevance \cite{lewis2020retrieval,sriramanan2024llm}. This approach is somewhat similar to the BBC's "At a Glance" summaries, which generate concise bullet-point overviews of lengthy government announcements through carefully crafted prompts and human editorial review before publication \cite{futureweek2025bbc,forbes2025bbc}. By constraining the LLM to summarize only verified government documents rather than generating novel content, KUBO would minimize hallucination risks while making lengthy posts and emergency alerts more accessible, particularly for low-literacy users.

Second, the SMS-based alert distribution would leverage the Philippines' existing cell broadcast infrastructure mandated by the Free Mobile Disaster Alerts Act (Republic Act 10639) \cite{ra10639} and NDRRMC systems \cite{dost2025alerto}. We can contextualize this technology also for non-emergency official announcements. This dual-channel approach, the mobile app plus the SMS feature, would ensure that residents without reliable internet access receive critical hyperlocal information.

Third, the community reporting feature would function similarly to Facebook to minimize user learning curves, but would implement identity verification through phone number and government ID registration to ensure credibility of users. Verified community members would be able to report local concerns (e.g., infrastructure damage, community hazards), which would be organized in a separate page from official announcements and could be upvoted or discussed through comments to enable peer validation of community concerns. This tiered verification system would enhance trust and reduce fraudulent content compared to anonymous community platforms.

Fourth, verified institutional posts by LGU officials, departments, and authorized government offices would be located on a dedicated page. Officials would manage their own posts directly to prevent account impersonation and to ensure that the verified news feed contains legitimate government communication. Managers of the posts are  appointed only through institutional email or credential authentication. This segregation of official content from community reporting creates a transparent institutional channel that clearly distinguishes government announcements from citizen input.

\section{Study Design}

\subsection{Participants}
We recruited 20 student participants, primarily from the University of the Philippines Los Baños. This demographic was chosen because students are highly engaged with digital platforms and strongly reliant on timely information. Participants were selected based on availability and willingness to participate; informed consent was obtained from all participants prior to testing.

\subsection{Experiment Setup}
To evaluate the effectiveness of our prototype, we compared Facebook, users' primary source of information, with KUBO. Our goal was to determine whether KUBO could significantly improve users' ability to access and comprehend crucial information. Two outcome measures guided the comparison: (1) the level of informedness, measured by participants' post-test quiz scores assessing awareness and recall, and (2) the time taken to retrieve information.

The independent variable in this study was the platform used (KUBO vs. Facebook), while the dependent variables were task completion time and quiz scores. We employed a within-subjects design so that each participant used both platforms under controlled conditions. To mitigate order effects, participants were randomly assigned to different sequences: half used KUBO first, while the others used Facebook first.

The following hypotheses were tested:

\subsubsection{Informedness}
\begin{itemize}
    \item \textit{Null hypothesis:} There is no significant difference in the level of informedness of participants when using the KUBO app compared to Facebook.
    \item \textit{Alternative hypothesis:} There is a significant difference between the level of informedness of participants when using the KUBO app compared to Facebook.
\end{itemize}

\subsubsection{Access Time}
\begin{itemize}
    \item \textit{Null hypothesis:} There is no significant difference in the time taken by participants to access crucial information when using the KUBO app compared to Facebook.
    \item \textit{Alternative hypothesis:} There is a significant difference in the time taken by participants to access crucial information when using the KUBO app compared to Facebook.
\end{itemize}

\subsection{Tasks and Procedure}
Participants completed a set of information-seeking tasks on both KUBO and Facebook, enabling a direct comparison of the two platforms. Tasks included:
\begin{itemize}
    \item Task 1: locating a recent post regarding a water outage,  
    \item Task 2: finding the official page of the UPLB Student Services Office (UPLB-SSO), and  
    \item Task 3: retrieving and filtering comments on a specific post.  
\end{itemize}

Each testing session began with a briefing. Participants were introduced to the KUBO prototype, its purpose, and its key features, including the Home page for verified updates and the Community page for user content. We also explained the overall testing procedure, the tasks to be performed, and the sequence of activities.

Participants were then asked to perform the tasks on both platforms. To control order effects, they were randomly assigned different sequences: some used Facebook first, while others started with KUBO. On each platform, the time taken to complete each task was recorded. After completing the first set of tasks, participants answered an information retrieval quiz. They then repeated the same set of tasks on the second platform, followed by a second quiz. 

Finally, participants filled out a post-task questionnaire assessing their overall experience, usability, and preferences. Each session concluded with a short debriefing where participants were thanked and the study goals were reiterated. The entire procedure lasted approximately 15 minutes per participant.

\subsection{Data Analysis}
To assess the effectiveness of our prototype, the researchers have analyzed quantitative data via task completion time, quiz scores, and participant feedback. We applied descriptive statistics to summarize data, followed by inferential tests to evaluate significance.

\subsubsection{Time Taken Per Task}
For task completion times, we first applied descriptive statistics and tested for normality using the Shapiro-Wilk test. Based on the results, we selected the appropriate inferential tests: Paired-samples t-tests for tasks with normally distributed differences, and Wilcoxon signed-rank tests for tasks that did not meet normality assumptions. We also conducted a Repeated Measures ANOVA to examine whether task order had any effect.

\subsubsection{Quiz Score Comparisons}
Quiz scores were similarly analyzed. We used descriptive statistics and the Shapiro-Wilk test to assess normality. Paired-samples t-tests were applied to compare scores between platforms, and Repeated Measures ANOVA was conducted to evaluate interactions between platform and presentation order, offering deeper insights into comprehension and information retention.

\subsubsection{Satisfaction and Preference}
Subjective data on satisfaction and preference were analyzed using descriptive statistics, given the paired and ordinal nature of the responses. We visualized mean user experience scores for both platforms using a bar graph.

\section{Results}

\subsection{Task Completion Time Across Platforms}

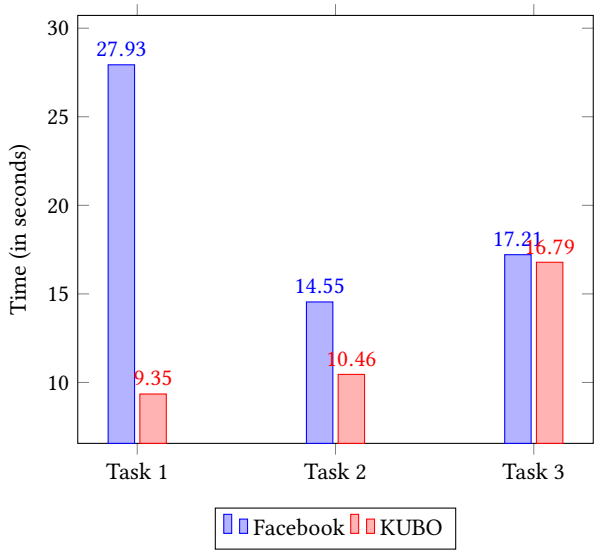
\begin{figure}[!htbp]
\centering
\begin{tikzpicture}
\begin{axis}[
    ybar,
    enlargelimits=0.15,
    legend style={at={(0.5,-0.15)}, anchor=north, legend columns=-1},
    ylabel={Time (in seconds)},
    symbolic x coords={Task 1, Task 2, Task 3},
    xtick=data,
    nodes near coords,
    nodes near coords align={vertical},
]
\addplot coordinates {(Task 1, 27.934) (Task 2, 14.545) (Task 3, 17.2135)};
\addplot coordinates {(Task 1, 9.3485) (Task 2, 10.4555) (Task 3, 16.7865)};
\legend{Facebook, KUBO}
\end{axis}
\end{tikzpicture}
\caption{Task Time Comparison (in seconds) for Facebook and KUBO}
\Description[Bar chart comparing task completion times]{A bar chart comparing average task completion times between Facebook and KUBO across three tasks. For Task 1, Facebook took 27.93 seconds while KUBO took about 9.35 seconds. For Task 2, Facebook took 14.55 seconds and KUBO 10.46 seconds. For Task 3, both platforms had similar times around 17 seconds. Overall, KUBO shows faster performance in Tasks 1 and 2.}
\label{fig:task-time}
\end{figure}

The primary aim of this test was to compare task completion times between the Facebook and KUBO. Participants were given three tasks and their time taken was recorded upon completion. Figure~\ref{fig:task-time} shows the summary of the comparison of the two platforms.

This Figure ~\ref{fig:task-time} shows a visual comparison of mean task completion times (in seconds) for Facebook and KUBO across three distinct tasks. An initial observation of the graph suggests that task completion times vary considerably, both between the two applications and across the different tasks (except on task 3 where both are close in the results).

There is a striking difference on Task 1. The bar representing Facebook indicates a substantially longer completion time compared to KUBO. This suggests that KUBO presents positive effects when it comes to searching for a specific post. However, we also need to account for the unfair comparison between a prototype and a real website application. There is a need for fetching data from the web, having irrelevant post presented, and the way a prototype is hard-coded for the task. The drastic difference may counter the difference by a small margin, suggesting that even with obvious bias between the platforms, KUBO is still the victorious platform for this task.

The next task proposes a slightly smaller difference between the two applications but suggesting that KUBO is still the one aligned with users when it comes to finding specific pages or people. The smaller gap could imply that Task 2 presents a scenario where both platforms perform more comparably, or that the specific advantages KUBO offered in Task 1 are less pertinent to Task 2. Additionally, the same concerns apply to this task about the unfairness of comparing the two platforms with very different implementations.

The final task about analyzing the details of a post and its interactions shows an almost similar mean score. Visually, there is little to no discernible difference between the two applications for this task. This suggests that for the activities involved in Task 3, neither application offered a clear efficiency advantage over the other.

To further investigate the observed differences in task completion times between the Facebook and KUBO applications, a series of statistical tests was conducted for each of the three tasks. The Shapiro-Wilk normality test suggests that the differences for Tasks 1 and 3 follow the normal distribution, while Task 2 does not. This suggests that for the Paired T-test, we use the Wilcoxon Signed-Ranks Test for Task 2 analysis. For this testing, alpha is set to 0.05. Table~\ref{tab:t-test} summarizes the results.

\begin{table}[!htbp]
  \caption{Test Results for Task Completion Times}
  \label{tab:t-test}
  \centering
  \begin{tabular}{l l l r r}
    \toprule
    \textbf{Pair} & \textbf{ } & \textbf{Test} & \textbf{Statistic} & \textbf{p} \\
    \midrule
    FB - Task 1 & K - Task 1 & Student's t & 6.026 & < 0.001 \\
    FB - Task 2 & K - Task 2 & Wilcoxon W & 134 & 0.294 \\
    FB - Task 3 & K - Task 3 & Student's t & 0.182 & 0.858 \\
    \bottomrule
  \end{tabular}
\end{table}

The Task 1 results indicated a statistically significant difference in completion times between Facebook and KUBO for Task 1, with p < 0.001. Examination of the means (as suggested by the positive t-value and the previously discussed graph) reveals that participants took significantly longer to complete Task 1 using Facebook compared to KUBO.

The Wilcoxon signed-rank test was used for the interpretation of Task 2 data. The test revealed no statistically significant difference in median completion times between Facebook and KUBO for Task 2, with p = 0.294. This means that Task 2 does not provide sufficient evidence that either of the platforms performs better than the other.

A paired-samples t-test for Task 3 data indicated no statistically significant difference in mean completion times between Facebook and KUBO for Task 3, with p = 0.858. Similar to Task 2, the data for this task does not provide sufficient evidence to conclude that either of the platforms performs better than the other.

\subsection{Quiz Score Comparisons Between the Two Platforms}

To understand whether the choice of application impacts users' ability to effectively gather and retain information, this study specifically investigated the level of informedness achieved by participants after using either the KUBO app or Facebook. This comparison aims to determine if one platform facilitates a greater degree of understanding or knowledge acquisition.

\begin{figure}[!htpb]
\centering
\begin{tikzpicture}
\begin{axis}[
    ybar stacked,
    width=\linewidth,
    height=7cm,
    bar width=6pt,
    ylabel={Score},
    xlabel={Participant},
    ymin=0, ymax=20,
    xtick={1,5,10,15,20},
    xticklabels={1,5,10,15,20},
    grid=major,
    enlarge x limits=0.03,
    legend style={at={(0.5,-0.2)}, anchor=north, legend columns=-1}
]

\addplot[fill=blue!50] coordinates {
    (1,4) (2,8) (3,3) (4,4) (5,5) (6,5) (7,7) (8,8) (9,7) (10,6)
    (11,6) (12,7) (13,6) (14,7) (15,6) (16,9) (17,7) (18,7) (19,8) (20,6)
};

\addplot[fill=red!60] coordinates {
    (1,9) (2,10) (3,7) (4,7) (5,8) (6,7) (7,9) (8,8) (9,9) (10,7)
    (11,7) (12,8) (13,10) (14,6) (15,6) (16,5) (17,9) (18,7) (19,9) (20,7)
};

\legend{Facebook Score, KUBO Score}
\end{axis}
\end{tikzpicture}
\caption{Stacked Bar Graph of Facebook and KUBO Scores by Participant}
\Description[Stacked bar graph comparing Facebook and KUBO scores]{A stacked bar graph showing Facebook scores as the lower segment and KUBO scores as the upper segment for each of twenty participants. For most participants, the KUBO segment is taller than the Facebook segment, and total bar heights highlight that KUBO generally outperforms Facebook in user scores.}
\label{fig:scores-graph}
\end{figure}
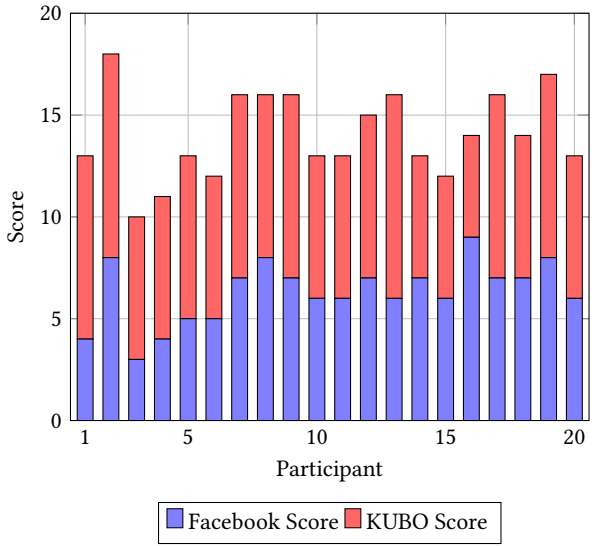

Visual examination of individual participant quiz scores, as seen in Figure~\ref{fig:scores-graph}, indicates high variability in informedness using Facebook compared to KUBO. Although neither app dominated the other for all individuals, KUBO scores (red line) often looked higher than or equal to Facebook scores (blue line), especially for participants who tried Facebook first. The lines intersect several times, which shows that the relative performance of the apps varied among subjects, with scores on both sites varying across the 0–10 scale.


Upon comparing the two presentation groups, a significant pattern can be observed. For participants who worked with Facebook initially (participants 1–10), there was a clearer tendency to score higher with KUBO. However, for participants who were exposed to KUBO initially (participants 11–20), results varied; while some continued to score higher with KUBO, a number achieved similar or even higher scores using Facebook. This visual disparity in score patterns between these two order groups implies that relative performance on the two applications may be affected by the order of exposure. These visible trends suggest that a straightforward across-samples comparison between Facebook and KUBO may be inadequate. The differing patterns across order groups strongly imply a possible interaction, in which the difference in informedness score between Facebook and KUBO may be contingent upon which application a participant accessed first. This visual proof supports the necessity for official statistical examination of whether these differences are significant and to formally test for an interaction effect.

After testing the normality of the difference in quiz scores, the Shapiro-Wilk Test found that the data are normally distributed. Hence, a normal Paired T-test is suitable to test our hypothesis. The result is shown in Table~\ref{tab:t-test2}.

\begin{table}[!htpb]
  \caption{Paired Samples T-Test Results for Quiz Scores}
  \label{tab:t-test2}
  \centering
  \begin{tabular}{l l l r r}
    \toprule
    \textbf{Pair} & \textbf{Test} & \textbf{Statistic} & \textbf{t} & \textbf{p} \\
    \midrule
    Facebook Score & KUBO Score & Student's t & -2.87 & 0.010 \\
    \bottomrule
  \end{tabular}
\end{table}

Findings indicate that the difference is statistically significant (p = 0.010 < 0.05). This finding implies the rejection of the null hypothesis, which was no difference in informedness. The negative t-statistic shows that participants scored significantly higher on average with the KUBO application than on Facebook. This result indicates a general benefit for KUBO in supporting user informedness under the conditions of the study. But since this particular test is giving an overall comparison, it does not control for the counterbalanced order with which participants accessed the apps. Thus, although it indicates a trend overall, it cannot determine whether such an advantage for KUBO existed consistently, whether Facebook or KUBO was presented first, or whether there was an interaction with presentation order.

\subsection{Satisfaction and Preference}
Beyond performance measures such as task completion time and informedness, we also examined participants' subjective experiences with both platforms. Satisfaction and preference ratings provide additional insight into how users perceived the usability and trustworthiness of KUBO compared to Facebook. These measures highlight not only which platform was more efficient, but also which one users found more intuitive and reliable for hyperlocal communication.

To evaluate users' subjective experience, we administered a post-task questionnaire for both platforms. All items were answered using a 5-point Likert scale (1 = Very Difficult / Very Dissatisfied / Very Ineffective, and 5 = Very Easy / Very Satisfied / Very Effective). The scale items were based on common usability constructs used in prior HCI evaluation (e.g., ease of use, satisfaction, and perceived effectiveness), and were adapted to the context of hyperlocal information-seeking.

The exact items were as follows:

\begin{itemize}
    \item \textbf{Ease of Use:} “How easy was it to navigate the platform to find the required information (e.g., locating a water outage post, finding the UPLB-SSO page, retrieving comments)?”
    \item \textbf{Satisfaction:} “How satisfied are you with the overall experience of finding information on the platform?”
    \item \textbf{Perceived Effectiveness:} “How effective was the platform in helping you access and understand crucial information quickly?”
\end{itemize}

\begin{table*}[!htpb]
  \caption{Repeated Measures ANOVA Within-Subjects Effects for Time Taken}
  \label{tab:anova}
  \centering
  \begin{tabular}{l r r r r r}
    \toprule
    \textbf{Effect} & \textbf{Sum of Squares} & \textbf{df} & \textbf{Mean Square} & \textbf{F} & \textbf{p} \\
    \midrule
    App Type & 1779.0 & 1 & 1779.0 & 16.423 & < 0.001 \\
    App Type $\times$ Order & 154.0 & 1 & 154.0 & 1.422 & 0.249 \\
    Residual & 1949.8 & 18 & 108.3 & & \\
    Task Number & 808.7 & 2 & 404.4 & 8.252 & 0.001 \\
    Task Number $\times$ Order & 932.6 & 2 & 466.3 & 9.517 & < 0.001 \\
    Residual & 1764.0 & 36 & 49.0 & & \\
    App Type $\times$ Task Number & 184.3 & 2 & 92.1 & 16.059 & < 0.001 \\
    App Type $\times$ Task Number $\times$ Order & 46.2 & 2 & 23.1 & 0.402 & 0.672 \\
    Residual & 2067.2 & 36 & 57.4 & & \\
    \bottomrule
  \end{tabular}
\end{table*}

Participants responded to the above items once for Facebook and once for KUBO. In addition, a set of forced-choice comparative items captured platform preference regarding:
\begin{itemize}
    \item Overall Preference (KUBO / Facebook / No Preference)
    \item Efficiency (which platform allowed faster task completion)
    \item Clarity of Information (which platform presented information more clearly)
    \item Credibility (which platform was more trustworthy)
\end{itemize}

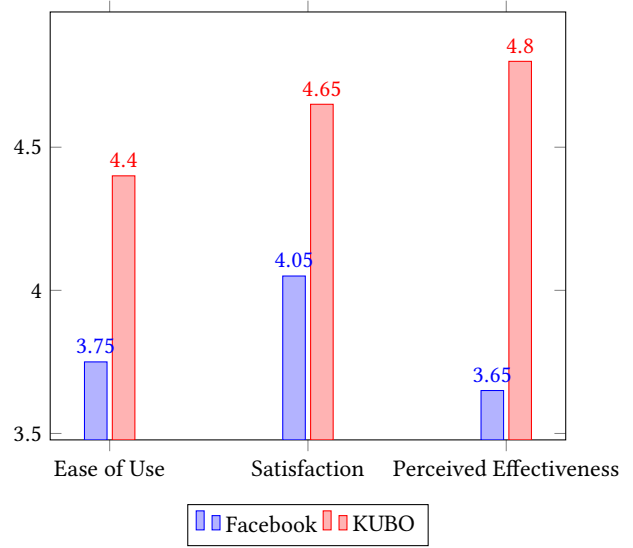
\begin{figure}[!htpb]
\centering
\begin{tikzpicture}
\begin{axis}[
    ybar,
    enlargelimits=0.15,
    legend style={at={(0.5,-0.15)}, anchor=north, legend columns=-1},
    symbolic x coords={Ease of Use, Satisfaction, Perceived Effectiveness},
    xtick=data,
    nodes near coords,
    nodes near coords align={vertical},
    bar width=0.3cm,
]
\addplot coordinates {(Ease of Use, 3.75) (Satisfaction, 4.05) (Perceived Effectiveness, 3.65)};
\addplot coordinates {(Ease of Use, 4.4) (Satisfaction, 4.65) (Perceived Effectiveness, 4.8)};
\legend{Facebook, KUBO}
\end{axis}
\end{tikzpicture}
\caption{Bar Graph of Mean User Experience Scores for Facebook and KUBO}
\Description[Bar chart comparing user experience ratings]{A bar chart comparing average user experience scores for Facebook and KUBO across three dimensions: ease of use, satisfaction, and perceived effectiveness. KUBO consistently scores higher than Facebook in all three categories, with mean ratings of approximately 4.4 for ease of use, 4.65 for satisfaction, and 4.8 for perceived effectiveness, compared to Facebook’s lower scores of 3.75, 4.05, and 3.65 respectively.}
\label{fig:scores-graph1}
\end{figure}

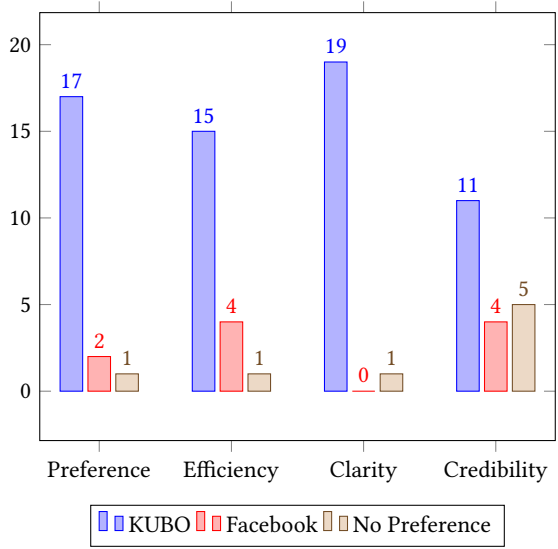
\begin{figure}[!htpb]
\centering
\begin{tikzpicture}
\begin{axis}[
    ybar,
    enlargelimits=0.15,
    legend style={at={(0.5,-0.15)}, anchor=north, legend columns=-1},
    symbolic x coords={Preference, Efficiency, Clarity, Credibility},
    xtick=data,
    nodes near coords,
    nodes near coords align={vertical},
    bar width=0.3cm,
]
\addplot coordinates {(Preference, 17) (Efficiency, 15) (Clarity, 19) (Credibility, 11)};
\addplot coordinates {(Preference, 2) (Efficiency, 4) (Clarity, 0) (Credibility, 4)};
\addplot coordinates {(Preference, 1) (Efficiency, 1) (Clarity, 1) (Credibility, 5)};
\legend{KUBO, Facebook, No Preference}
\end{axis}
\end{tikzpicture}
\caption{Bar Graph of Preference and Evaluation Metrics}
\Description[Bar chart comparing user preference and evaluation criteria]{A bar chart comparing user preferences and evaluation metrics for KUBO and Facebook across four categories: preference, efficiency, clarity, and credibility. Seventeen participants preferred KUBO versus two for Facebook and one with no preference. For efficiency, KUBO received fifteen votes compared to four for Facebook and one undecided. For clarity, KUBO scored highest with nineteen votes while Facebook had none and one participant expressed no preference. For credibility, eleven participants chose KUBO, four chose Facebook, and five had no preference. Overall, KUBO was consistently favored across all evaluation metrics.}
\label{fig:scores-graph2}
\end{figure}

These items provide a structured basis for the subjective evaluations reported in Figure~\ref{fig:scores-graph1} and Figure~\ref{fig:scores-graph2}.

Simply observing Figure~\ref{fig:scores-graph1}, which plots mean user experience scores, shows a uniform trend in favor of KUBO over Facebook for all three constructs measured: Ease of Use, Satisfaction, and Perceived Effectiveness. For every metric, the bar for KUBO is taller to the eye, indicating that participants, in general, indicated a more favorable experience for KUBO. This trend suggests a typical user preference for KUBO under these subjective ratings.

In particular, KUBO is seemingly rated significantly higher than Facebook for both Ease of Use and Satisfaction. The most significant visual contrast appears in Perceived Effectiveness, in which KUBO got significantly higher scores than Facebook, suggesting participants believed KUBO was dramatically more useful for their work. 

Looking at Figure~\ref{fig:scores-graph2}, the graphical depiction of participant preference for four factors—Preference, Efficiency, Clarity of Information, and Credibility—demonstrates an overwhelming bias towards the KUBO prototype compared to Facebook. In every factor, the blue bar for KUBO stands much higher than the red bars for Facebook or the yellow bars for No Preference, indicating that KUBO was overall rated higher. This trend is most dramatic for Clarity of Information, where KUBO was selected by virtually all participants, and also extremely clear for Overall Preference and Efficiency, where KUBO was selected with a strong majority.

Whereas KUBO continually appeared as the preferred choice, the intensity of this favoring differed slightly across the measures. The favoring of KUBO was strongest in terms of Clarity of Information and remained extremely high for Overall Preference and Efficiency. For Credibility, KUBO remained the most chosen alternative, but the lead was narrower, with Facebook being more picked and having a higher percentage of participants who responded "No Preference" than to the other criteria. This indicates that while KUBO overall was viewed as better, views of credibility were more polarized or in doubt among the participants.

\subsection{Order Effects Analysis}

To determine whether the sequence in which participants used the two platforms influenced performance, we examined order effects for both task completion time and informedness scores using repeated-measures ANOVA.

\subsubsection{Task Completion Time}

Task completion time showed a significant main effect of App Type ($p < 0.001$), indicating that participants generally completed tasks faster in KUBO than in Facebook. However, the App Type $\times$ Order interaction was not significant ($p = 0.249$), meaning this speed advantage remained consistent regardless of whether participants used KUBO first or Facebook first. This indicates that the efficiency benefits of KUBO were not influenced by learning or familiarity effects introduced by task sequence.

\subsubsection{Informedness Scores}

In contrast, informedness (measured by post-task quiz scores) demonstrated a significant App Type $\times$ Order interaction ($p = 0.023$). Participants who used KUBO first scored higher overall, while those who used Facebook first showed smaller differences between the two platforms. This suggests that starting with KUBO helped participants build a clearer mental model of the information, which persisted even when switching to Facebook.

\begin{table}[!htpb]
  \caption{Repeated Measures ANOVA for Quiz Score}
  \label{tab:anova-quiz}
  \centering
  \begin{tabular}{l r r r r r}
    \toprule
    \textbf{Effect} & \textbf{SS} & \textbf{df} & \textbf{MS} & \textbf{F} & \textbf{p} \\
    \midrule
    App Type & 23.4 & 1 & 23.4 & 19.005 & < 0.001 \\
    App Type $\times$ Order & 3.1 & 1 & 3.1 & 6.056 & 0.023 \\
    Residual & 22.1 & 18 & 1.23 & & \\
    \bottomrule
  \end{tabular}
\end{table}

In summary, order did not influence task efficiency, but it did influence comprehension. Beginning with KUBO appears to establish a structured and clearer understanding of the information environment, whereas beginning with Facebook may require additional effort to interpret content before clarity is achieved. This pattern is reflected in participants' comments describing KUBO as “straightforward” and Facebook as requiring “extra steps to confirm details.”

\section{Discussion}
The evaluation results provide evidence that, in general, KUBO better supports hyperlocal information-seeking compared to Facebook. Among the three experimental tasks, only Task 1 (i.e., locating a recent post about a water outage) showed a statistically significant difference; that means that participants completed it more quickly in KUBO. This result highlights the strength of KUBO's design in surfacing urgent and time-sensitive information. The verified advisories and a structured feed reduced the effort typically required to search across multiple pages or posts on social media. The finding aligns with one of the core user needs identified in our user research (i.e., the rapid access to relevant alerts) and demonstrates how hyperlocal systems can directly support informedness in critical situations such as outages or emergencies.

In contrast, Task 2 (i.e., finding the official UPLB-SSO page) and Task 3 (i.e., retrieving and filtering comments) did not yield significant differences between the two platforms. One explanation is that both platforms provide relatively straightforward mechanisms for page discovery and comment browsing, and these reduced the margin for KUBO to outperform Facebook. Another possibility is that these tasks rely less on urgency-sensitive features and more on general navigation; the familiarity with Facebook may have offset KUBO's design advantages.

KUBO also demonstrated clear improvements in informedness as measured by quiz performance. Participants scored significantly higher after using KUBO. This result suggests that the platform not only facilitated faster access but also enhanced comprehension and recall of information. The reduced cognitive load and ambiguity when processing information could be likely explained by KUBO's emphasis on verified, LGU-sourced advisories and concise summaries

Note that the specific effectiveness and accuracy of the AI-generated summaries were not evaluated in this study. Our current design does not yet address the significant risks associated with this feature, such as the potential for AI-driven factual inaccuracies (hallucinations), the propagation of bias, or how such errors could negatively affect the platform's core goal of enhancing information credibility. We hypothesize that this feature would address information overload, but its actual impact on user trust remains a critical assumption that must still be tested. In contrast, Facebook's engagement-driven feed often required users to filter through irrelevant or redundant posts, making it more difficult to identify and retain key details. These findings align with our user research themes, particularly the need for trusted sources, relevant alerts, and reduced search effort, and demonstrate how hyperlocal communication platforms can support not only faster access but also deeper comprehension of critical community updates.

On subjective measures, KUBO consistently outperformed Facebook in user ratings of ease of use, satisfaction, and perceived effectiveness, with mean scores at the top of the scale. Preference data showed similar trends: the majority of participants favored KUBO for overall use, efficiency, and especially clarity of information, where 19 of 20 participants chose KUBO over Facebook. However, it is important to note that all participants were university students who are digitally native and familiar with mobile applications. Their familiarity with structured information systems and institutional communication channels may have influenced the degree to which KUBO's interface and content presentation felt intuitive and trustworthy. Usage patterns and preferences may differ among community members with varying levels of digital literacy or different expectations of online civic communication platforms. In terms of credibility, KUBO was also preferred, but a notable portion of participants remained neutral. This suggests that perceptions of credibility may depend not only on interface design but also on broader trust in institutions and sources.

\section{Design Implications}

Our findings point to several design implications for hyperlocal communication platforms and digital news systems. The significantly reduced task completion times and higher information recall observed with KUBO suggest that efficiency and clarity are crucial drivers of informedness. In contrast to mainstream social media, hyperlocal platforms must foreground verified and locally relevant content. Features such as attribution of posts to barangay offices, local government units (LGUs), or accredited organizations, as well as visible verification markers, can strengthen trust and, at the same time, reduce the risks of misinformation.

The results highlight the importance of urgency- and location-sensitive content curation. Participants rated KUBO highly because of the streamlined access to geographically proximal and temporally urgent updates improves user experience. This, in turn, improved efficiency and clarity of information. Hyperlocal systems should prioritize relevance to place and time, such as geolocation filters, neighborhood-based channels, and ranking mechanisms that elevate emergency advisories above routine content, unlike algorithmic curation in social media.

High user ratings for credibility and preference highlight the need for inclusive and participatory design. Hyperlocal systems must balance multiple stakeholder roles while maintaining accessible, low-bandwidth interfaces that accommodate varying levels of digital literacy. Providing differentiated channels for official announcements and community reporting enables broad participation, and at the same time, preserves clarity and trust.

Ultimately, the study suggests that hyperlocal communication is most effective when it leverages existing cultural and social practices. Even as digital interfaces enabled faster completion of tasks and more accurate recall, participants still valued elements of word-of-mouth and community-based trust networks. Incorporating features such as neighbor recommendations, group-shared posts, or visible community validations can extend these offline practices into digital spaces. When combined with institutional verification, such mechanisms help bridge the gap between informal trust cues and formal sources of information.

\section{Conclusion}
This work presented KUBO (Kumunidad at Balitang Opisyal), a dual-channel platform for hyperlocal communication. Our user-centered design process through contextual inquiry and semi-structured interviews has surfaced five recurring needs: (1) rapid delivery of urgent information, (2) trust through verified sources and clear attribution, (3) mobile-first access with SMS fallbacks for low connectivity, (4) reduced effort via concise, relevant updates, and (5) support for resident participation. Affinity diagramming produced five high-level themes, along with the user personas, directly informed KUBO's core features: a Home feed for verified advisories, a Community space for resident reports, verification badges, AI-generated summaries, and SMS alerts. Iteration from low- to high-fidelity prototypes refined the information architecture, filtering, and follow/report mechanisms.

Our within-subjects evaluation with 20 participants shows that KUBO improves efficiency and informedness and is preferred by users over Facebook. For task completion time, KUBO was significantly faster on the specific post-finding task (Task 1; $p<0.001$). For informedness, participants scored significantly higher with KUBO (paired $t$-test, $p=0.010$). Subjective measures consistently favored KUBO across ease of use, satisfaction, perceived effectiveness, and overall preference.

These results suggest that combining verified, LGU-sourced updates with community reporting and multi-channel delivery can meaningfully reduce time-to-information and increase comprehension in hyperlocal contexts. The implications for design extend beyond KUBO: hyperlocal communication platforms and news systems should prioritize verified and attributed sources to address the credibility gap of social media, implement urgency- and location-sensitive ranking to ensure rapid access to critical information, and support lightweight, participatory features that reflect community trust practices. In doing so, platforms can bridge the gap between institutional reliability and everyday neighborhood communication. These offer an alternative model to engagement-driven social media feeds.

Like most early-stage design research, this study carries certain limitations. First, the evaluation was conducted with a high-fidelity prototype rather than a fully deployed system. This allowed us to isolate and test the core design features, but the production deployments will inevitably introduce factors such as real-world adoption dynamics that could affect performance and usability. 

Second, our participant pool was relatively small and skewed toward university students. This limits the generalizability of our findings to the broader population. Community members from diverse demographics, including elderly residents, barangay officials, and small business owners, may have different expectations, literacy levels, and trust thresholds that future studies should address. 

Third, while our design included an AI summarization feature to address user needs for concise updates, this study did not measure its specific effectiveness or its impact on user trust. The concerns of AI reliability, such as the potential for hallucinations or bias, could affect the platform's goal of trustworthiness. Future work must therefore not only measure the feature's utility but also evaluate the content's safety and effect on perceived credibility before deployment.

Future work will therefore involve piloting KUBO across multiple barangays and community types to test its robustness in broader and valid settings. Integration with LGU information systems will be critical for ensuring continuity of verified updates and for embedding the platform within existing governance workflows. Longitudinal studies will also allow us to assess how hyperlocal platforms shape community resilience, trust dynamics, and resistance to misinformation over time. 

Ultimately, the design insights from KUBO contributed to an understanding of hyperlocal digital systems, demonstrating how sociotechnical interventions can foster participatory governance, strengthen information integrity, and support sustainable community development.

\begin{acks}
J.I.N. Gojo Cruz gratefully acknowledges the support provided by the University of the Philippines Los Baños through its Academic Development Fund.
\end{acks}

\bibliographystyle{ACM-Reference-Format}
\bibliography{sample-base}

@mastersthesis{afable2020facebook,
  author = {Afable, Nicole Marie D.},
  title        = {Using Facebook for public engagement: An analysis of the public Facebook pages of the local government units in Metro Manila},
  school       = {De La Salle University},
  year         = {2020},
  url          = {https://animorepository.dlsu.edu.ph/etd_masteral/5935},
  type         = {Master's thesis}
}

@misc{meltwater2024social,
  author       = {Meltwater},
  title        = {Social Media Statistics in the Philippines [Updated 2025]},
  year         = {2025},
  url          = {https://www.meltwater.com/en/blog/social-media-statistics-philippines},
  note         = {Accessed: 2025-05-18}
}

@misc{stephens2018risk,
  author       = {Stephens, Sonia},
  title        = {Building Better Tools for Risk Communication with User-Centered Design},
  year         = {2018},
  howpublished = {AGU Blogosphere},
  url          = {https://blogs.agu.org/sciencecommunication/2018/04/23/building-better-tools-for-risk-communication-with-user-centered-design},
  note         = {Accessed: 2025-05-18}
}

@inproceedings{pedrosa2022user,
  title={Applying user-centered design on digital transformation of public services: A case study in brazil},
  author={Pedrosa, Glauco Vitor and Judice, Andrea and Judice, Marcelo and Ara{\'u}jo, Leonardo and Fleury, Fabiola and Figueiredo, Rejane},
  booktitle={Proceedings of the 23rd Annual International Conference on Digital Government Research},
  pages={372--379},
  year={2022}
}

@misc{pia2020local,
  author       = {{Philippine Information Agency}},
  title        = {Local Government News},
  year         = {2020},
  url          = {https://pia.gov.ph/about/},
  note         = {Accessed: 2025-05-18}
}

@inproceedings{firdaus2016qlue,
  author       = {Firdaus, Muhamad Shendy Adam and Irwansyah, Irwansyah and Djaja, Komara},
  title        = {Mobile Apps as Government Communication Media in Urban Public Services: Case Study – The Usage of Qlue Application by Jakarta Provincial Government},
  booktitle    = {Sustainable Development and Planning VIII},
  pages        = {417--430},
  year         = {2016},
  doi          = {10.2495/SDP160351}
}

@misc{wardle2017information,
  author       = {Wardle, Claire and Derakhshan, Hossein},
  title        = {Information Disorder: Toward an Interdisciplinary Framework for Research and Policymaking},
  year         = {2017},
  institution  = {Council of Europe},
  note         = {Accessed: 2025-05-18}
}

@misc{beltran2022tsunami,
  author       = {Beltran, Michael},
  title        = {Disinformation reigns in Philippines as Marcos Jr takes top job},
  year         = {2022},
  howpublished = {Al Jazeera},
  note         = {Accessed: 2025-05-18}
}

@misc{rappler2023disaster,
  author       = {Pasion, Lorenz},
  title        = {Disaster-related Lies, Disinformation Debunked by Rappler in 2023},
  year         = {2023},
  url          = {https://www.rappler.com/environment/disasters/lies-disinformation-related-disasters-debunked-2023/},
  note         = {Accessed: 2025-05-18}
}

@article{fabella2022fakenews,
  author       = {Fabella, Frederick Edward T.},
  title        = {Investigating Factors that Influence the Belief in and Sharing of Social Media News as well as the Attitudes toward Fake News of Selected Filipinos},
  journal      = {Cognizance Journal of Multidisciplinary Studies},
  volume       = {2},
  number       = {5},
  pages        = {1--12},
  year         = {2022},
  url          = {https://www.researchgate.net/publication/360845646_Investigating_Factors_that_Influence_the_Belief_in_and_Sharing_of_Social_Media_News_as_well_as_the_Attitudes_toward_Fake_News_of_Selected_Filipinos},
  note         = {Accessed: 2025-05-18}
}

@article{bakshy2015exposure,
  author       = {Bakshy, Eytan and Messing, Solomon and Adamic, Lada A.},
  title        = {Exposure to Ideologically Diverse News and Opinion on Facebook},
  journal      = {Science},
  volume       = {348},
  number       = {6239},
  pages        = {1130--1132},
  year         = {2015},
  url          = {https://www.science.org/doi/10.1126/science.aaa1160},
  doi          = {10.1126/science.aaa1160},
  note         = {Accessed: 2025-05-18}
}

@book{pariser2011filterbubble,
  author       = {Pariser, Eli},
  title        = {The Filter Bubble: How the New Personalized Web Is Changing What We Read and How We Think},
  publisher    = {Penguin Books},
  year         = {2012},
  note         = {Accessed: 2025-05-18}
}

@article{metzler2024social,
  title={Social drivers and algorithmic mechanisms on digital media},
  author={Metzler, Hannah and Garcia, David},
  journal={Perspectives on Psychological Science},
  volume={19},
  number={5},
  pages={735--748},
  year={2024},
  publisher={Sage Publications Sage CA: Los Angeles, CA}
}

@article{hu2021shedr,
  title={SHEDR: an end-to-end deep neural event detection and recommendation framework for hyperlocal news using social media},
  author={Hu, Yuheng and Hong, Yili},
  journal={INFORMS Journal on Computing},
  volume={34},
  number={2},
  pages={790--806},
  year={2021},
  publisher={INFORMS}
}

@misc{meese2019bloom,
  author       = {Meese, Emma},
  title        = {Video Tutorial: Local search and analytics for hyperlocal publishers using Bloom},
  year         = {2019},
  howpublished = {Independent Community News Network, Cardiff University},
  url          = {https://www.communityjournalism.co.uk/resources/video-tutorial-local-search-and-analytics-for-hyperlocal-publishers-using-bloom/},
  note         = {Accessed: 2025-05-18}
}

@article{hartwig2024landscape,
  author    = {Hartwig, Katrin and Doell, Frederic and Reuter, Christian},
  title     = {The Landscape of User-centered Misinformation Interventions -- A Systematic Literature Review},
  journal   = {ACM Computing Surveys},
  volume    = {56},
  number    = {11},
  pages     = {292:1--292:36},
  year      = {2024},
  month     = {November},
  publisher = {Association for Computing Machinery},
  address   = {New York, NY, USA},
  doi       = {10.1145/3674724},
  url       = {https://doi.org/10.1145/3674724}
}

@article{malhotra2023user,
  title={User experiences and needs when responding to misinformation on social media},
  author={Malhotra, Pranav and Zhong, Ruican and Kuan, Victor and Panatula, Gargi and Weng, Michelle and Bras, Andrea and Sehat, Connie Moon and Roesner, Franziska and Zhang, Amy},
  journal={Harvard Kennedy School Misinformation Review},
  year={2023},
  url={https://doi.org/10.37016/mr-2020-129}
}

@article{kavanaugh2012hyper,
    title = {(Hyper) local news aggregation: Designing for social affordances},
    journal = {Government Information Quarterly},
    volume = {31},
    number = {1},
    pages = {30-41},
    year = {2014},
    issn = {0740-624X},
    doi = {https://doi.org/10.1016/j.giq.2013.04.004},
    url = {https://www.sciencedirect.com/science/article/pii/S0740624X13001251},
    author = {Andrea Kavanaugh and Ankit Ahuja and Samah Gad and Sloane Neidig and Manuel A. Pérez-Quiñones and Naren Ramakrishnan and John Tedesco}
}

@article{gulyas2024three,
  title={The three “Cs” of digital local journalism: community, Commitment and continuity},
  author={Gulyas, Agnes and Hess, Kristy},
  journal={Digital Journalism},
  volume={12},
  number={1},
  pages={6--12},
  year={2024},
  publisher={Taylor \& Francis}
}

@article{noman2024designing,
  title={Designing social media to foster user engagement in challenging misinformation: a cross-cultural comparison between the UK and Arab countries},
  author={Noman, Muaadh and Gurgun, Selin and Phalp, Keith and Ali, Raian},
  journal={Humanities and Social Sciences Communications},
  volume={11},
  number={1},
  pages={1--13},
  year={2024},
  publisher={Nature}
}

@article{gurgun2024motivated,
  title={Motivated by Design: A Codesign Study to Promote Challenging Misinformation on Social Media},
  author={Gurgun, Selin and Arden-Close, Emily and Phalp, Keith and Ali, Raian},
  journal={Human Behavior and Emerging Technologies},
  volume={2024},
  number={1},
  pages={5595339},
  year={2024},
  publisher={Wiley Online Library}
}

@techreport{designcouncil2005double,
  title={The Double Diamond: A universally accepted depiction of the design process},
  author={{Design Council}},
  year={2004},
  institution={Design Council, UK},
  url={https://www.designcouncil.org.uk/our-resources/framework-for-innovation/},
  note         = {Accessed: 2025-05-18}
}

@article{brown2009change,
  title={Change by design: How design thinking transforms organizations and inspires innovation},
  author={Brown, Tim},
  journal={Harper Business},
  year={2009}
}

@book{holtzblatt1997contextual,
  title={Contextual design: defining customer-centered systems},
  author={Holtzblatt, Karen and Beyer, Hugh},
  year={1997},
  publisher={Morgan Kaufmann Publishers Inc.},
  isbn={978-0-08-050304-2}
}

@inproceedings{lucero2015using,
  title={Using affinity diagrams to evaluate interactive prototypes},
  author={Lucero, Andr{\'e}s},
  booktitle={IFIP conference on human-computer interaction},
  pages={231--248},
  year={2015},
  organization={Springer}
}

@book{nielsen2013personas,
  title={Personas - user focused design},
  author={Nielsen, Lene},
  year={2019},
  publisher={Springer},
  isbn={978-1-4471-7427-1},
  doi={https://doi.org/10.1007/978-1-4471-7427-1}
}

@article{nielsen1995conduct,
  title={How to conduct a heuristic evaluation},
  author={Moran, Kate and Gordon, Kelley},
  journal={Nielsen Norman Group},
  year={2023}
}

@misc{egovph2022app,
  title={eGov PH: Elevating convenience to government experience},
  author={{Department of Information and Communications Technology}},
  year={2022},
  howpublished={eGov PH Official Website},
  url={https://e.gov.ph}
}

@misc{pco2025egovph,
  title={eGov PH app has more than a thousand government services},
  author={{Presidential Communications Office}},
  year={2025},
  month={June},
  url={https://pco.gov.ph/news_releases/egov-ph-app-has-more-than-a-thousand-government-systems-integrated-dict/}
}

@misc{mandaluyong2022namayan,
  title={First barangay mobile app launched in Namayan, Mandaluyong},
  author={Herrera, Jenny},
  year={2022},
  month={September},
  howpublished={Mandaluyong City Government Website},
  url={https://mandaluyong.gov.ph/first-barangay-mobile-app-launched-in-namayan-mandaluyong-city-digitization-of-basic-barangay-services-made-possible-through-namayan-mobile-app/},
  note         = {Accessed: 2025-05-18}
}

@misc{naga2025mynaga,
  title={Now available: MyNaga app - Naga City at your fingertips},
  author={{City Government of Naga}},
  year={2025},
  howpublished={Official Naga City Website},
  url={https://www2.naga.gov.ph/available-mynaga-app/}
}

@misc{playgoogle2025hope,
  title={1Hope: Community mobile app for emergency reporting},
  author={{Barangay Bagong Pag-asa QC}},
  year={2025},
  howpublished={Barangay Bagong Pag-asa QC Website},
  url={https://bagongpagasaqc.defensys.ph/}
}

@misc{ra10639,
  title = {{Republic Act No. 10639: An Act Mandating the Telecommunications Service Providers to Send Free Mobile Alerts in the Event of Natural and Man-made Disasters and Calamities}},
  author = {{Congress of the Philippines}},
  year = {2014},
  url = {https://ldr.senate.gov.ph/sites/default/files/2023-02/ra%252010639.pdf}
}

@article{dost2025alerto,
  title={How 'Alerto PH' app can boost LGUs' emergency \& disaster response},
  author={Mondares, Claire Bernadette A.},
  year={2025},
  month={July},
  url={https://www.dost.gov.ph/knowledge-resources/news/86-2025-news/4098-how-alerto-ph-app-can-boost-lgus-emergency-disaster-response-capabilities-the-rise-of-filipino-inventors-in-drrm.html},
  journal={Department of Science and Technology}
}

@misc{playgoogle2025jodel,
  title={Jodel: Hyperlocal community app},
  author={{The Jodel Venture GmbH}},
  year={2025},
  howpublished={Google Play Store},
  url={https://play.google.com/store/apps/details?id=com.tellm.android.app&hl=en}
}

@misc{neighbrsnook2025benefits,
  title={The benefits of hyperlocal social networks},
  author={{Neighbrsnook}},
  year={2024},
  url={https://neighbrsnook.com/blogs/benefits-of-hyperlocal-social-networks/#about}
}

@article{theverge2024particle,
  title={Particle is a new app using AI to organize and summarize the news},
  author={Pierce, David},
  journal={The Verge},
  year={2024},
  month={November},
  url={https://www.theverge.com/2024/11/12/24293993/particle-news-app-ai-summaries}
}

@misc{playgoogle2024letmeknow,
  title={AI news summaries: LetMeKnow},
  author={{Diemz}},
  year={2024},
  howpublished={Google Play Store},
  url={https://play.google.com/store/apps/details?id=com.diemz.letmeknow&hl=en}
}

@article{statista2025socmed,
  title={Social media in the Philippines - statistics \& facts},
  journal={Statista Research Department},
  author={Balita, Christy},
  year={2025},
  url={https://www.statista.com/topics/5914/social-media-usage-in-the-philippines/}
}

@online{spiralytics2025facts,
  title   = {Understanding Social Media in the Philippines: 2025 Facts \& Statistics},
  author={Caparas, Jozella},
  year={2025},
  url={https://www.spiralytics.com/blog/social-media-in-the-philippines-facts-and-statistics/},
  organization = {Spiralytics}
}

@article{delacruz2021surfing,
  title={Surfing the waves of infodemics: Building a cohesive Philippine framework against misinformation},
  author={Dela Cruz, Joseph Rem and Tiu, Christl Jan and Velasco, Raphael Ian and Abella, Hannah Joyce and Dela Cruz, Mark Vincent and Lacson, Jemil Austin and Macatangay, Ian Oliver and Ong, Erika and Paguntalan, Jaypee and Doroja, Nanette},
  journal={Journal of Asian Medical Students' Association},
  volume={9},
  number={1},
  year={2021}
}

@article{pennycook2021shifting,
  title={Shifting attention to accuracy can reduce misinformation online},
  author={Pennycook, Gordon and Epstein, Ziv and Mosleh, Mohsen and Arechar, Antonio A. and Eckles, Dean and Rand, David G.},
  journal={Nature},
  volume={592},
  number={7855},
  pages={590--595},
  year={2021},
  publisher={Nature Publishing Group UK London}
}

@article{bhuiyan2021nudgecred,
  title={NudgeCred: Supporting news credibility assessment on social media through nudges},
  author={Bhuiyan, Md Momen and Horning, Michael and Lee, Sang Won and Mitra, Tanushree},
  journal={Proceedings of the ACM on Human-Computer Interaction},
  volume={5},
  number={CSCW2},
  pages={1--30},
  year={2021},
  publisher={ACM New York, NY, USA}
}

@article{porter2021global,
  title={The global effectiveness of fact-checking: Evidence from simultaneous experiments in Argentina, Nigeria, South Africa, and the United Kingdom},
  author={Porter, Ethan and Wood, Thomas J.},
  journal={Proceedings of the National Academy of Sciences},
  volume={118},
  number={37},
  pages={e2104235118},
  year={2021},
  publisher={National Academy of Sciences}
}

@inproceedings{lewis2020retrieval,
  author = {Lewis, Patrick and Perez, Ethan and Piktus, Aleksandra and Petroni, Fabio and Karpukhin, Vladimir and Goyal, Naman and K\"{u}ttler, Heinrich and Lewis, Mike and Yih, Wen-tau and Rockt\"{a}schel, Tim and Riedel, Sebastian and Kiela, Douwe},
    title = {Retrieval-augmented generation for knowledge-intensive NLP tasks},
    year = {2020},
    isbn = {9781713829546},
    publisher = {Curran Associates Inc.},
    address = {Red Hook, NY, USA},
    booktitle = {Proceedings of the 34th International Conference on Neural Information Processing Systems},
    articleno = {793},
    numpages = {16},
    location = {Vancouver, BC, Canada},
    series = {NIPS '20}
}

@article{sriramanan2024llm,
  title={LLM-Check: Investigating Detection of Hallucinations in Large Language Models},
  author={Sriramanan, Gaurang and Bharti, Siddhant and Sadasivan, Vinu Sankar and Saha, Shoumik and Kattakinda, Priyatham and Feizi, Soheil},
  journal={Advances in Neural Information Processing Systems},
  volume={37},
  pages={34188--34216},
  year={2024}
}

@article{futureweek2025bbc,
  title={BBC introduces gen AI summary in news stories},
  author={Duffy, Katie},
  year={2025},
  month={June},
  url={https://futureweek.com/bbc-introduces-gen-ai-summary-in-news-stories/},
  journal={Futureweek}
}

@article{forbes2025bbc,
  title={BBC rolls out AI summaries and style tool in newsroom pilot},
  author={Schmelzer, Ron},
  year={2025},
  month={June},
  url={https://www.forbes.com/sites/ronschmelzer/2025/06/29/bbc-rolls-out-ai-summaries-and-style-tool-in-newsroom-test/},
  journal={Forbes}
}

@article{dow2005wizard,
  title={Wizard of Oz support throughout an iterative design process},
  author={Dow, Steven and MacIntyre, Blair and Lee, Jaemin and Oezbek, Christopher and Bolter, Jay David and Gandy, Maribeth},
  journal={IEEE Pervasive Computing},
  volume={4},
  number={4},
  pages={18--26},
  year={2005},
  publisher={IEEE}
}

\end{document}